# Green laser powder bed fusion based fabrication and rate-dependent mechanical properties of copper lattices

Sung-Gyu Kang [a*], Ramil Gainov [b], Daniel Heußen [c], Sören Bieler [d], Zhongji Sun [e], Kerstin Weinberg [d], Gerhard Dehm [a], and Rajaprakash Ramachandramoorthy [a*]

[a] Max-Planck-Institut für Eisenforschung GmbH, Max-Planck-Straße 1, 40237 Düsseldorf, Germany

[b] Institute of Mineral Resources Engineering, RWTH Aachen University, Wüllnerstraße 2, 52062 Aachen, Germany

[c] Fraunhofer-Institut für Lasertechnik ILT, Steinbachstraße. 15, 52074 Aachen, Germany

[d] Lehrstuhl für Festkörpermechanik, Universität Siegen, Paul-Bonatz-Straße 9-11, 57068 Siegen, Germany

[e] Institute of Materials Research and Engineering, Fusionopolis Way 2, 138634 Singapore

*Corresponding author: Dr. Sung-Gyu Kang and Dr. Rajaprakash Ramachandramoorthy

E-mail: s.kang@mpie.de and r.ram@mpie.de

# Abstract

Lattice structures composed of periodic solid frames and pores can be utilized in energy absorption applications due to their high specific strength and large deformation. However, these structures typically suffer from post-yield softenings originating from the limited plasticity of available material choices. This study aims to resolve such an issue by fabricating lattice structures made of ductile pure copper (Cu). For the first time, Cu lattice structures are fabricated through laser-powder bed fusion (L-PBF) with green laser ($\lambda = 515$ nm). Structural and microstructural analysis confirm that the lattice structures consist of well-defined unit-cells and show dense microstructure. The deformation behavior is investigated under a wide range of strain rates from ~0.001 /s to ~1000 /s. The stress-strain curves exhibit a smooth and continuous deformation without any post-yield softening, which can be attributed to the intrinsic mechanical properties of Cu. Correlated with post-mortem microscopy examination, the rate-dependent deformation behavior of pure Cu lattice structures is investigated and rationalized. The current work suggests that the complex Cu architectures can be fabricated by L-PBF with green laser and the lattice structures made of ductile metal are suitable for dynamic loading applications.

# 1. Introduction

Lattice structures are composed of periodic solid frames and pores. Its periodicity enables higher specific strength compared to unstructured foams Combined with large deformation until densification under constant loads arising from the low relative density, the lattice structures can exhibit a larger absorption capacity compared to solid materials under compression[1–3]. Recent advances in the additive manufacturing (AM) processes have enabled several studies on lattice structures with envisioned applications in a variety of sectors including aerospace [4,5], biomaterials [6–8], mechanical band gap engineering [9–11], and impact absorption [12–14]. The deformation characteristics of lattice structures strongly depend on the geometry and the material. From the geometrical point of view, the strength of the lattices can be tuned by controlling the orientation of load-bearing solid frames, density, and connectivity. To date, various lattice geometries such as open-cell truss lattices [12,14–16], closed-cell plate lattices [17–19], triply periodic minimal surface structures [20–22], and shell structures [23,24], have been investigated both experimentally and computationally. From the material point of view, previous studies have shown that the intrinsic plasticity of the material strongly affects the deformation characteristics of lattice structures. The microlattice structures made of brittle materials such as alumina [24–26] and pyrolytic carbon [17] show high strength but low plasticity. On the other hand, the macroscale lattice structures made of Ti- [27,28], Al- [29], and Fe-alloys [30] show continuous deformation with large plastic strains, leading to superior energy absorption capacity as compared to random foam structures. Interestingly, even in these lattice structures made of structural load-bearing alloy systems, abrupt softening may occur during deformation, which degrades their energy absorption capacity [12]. This abrupt softening behavior originates from the buckling or cracking of the strut structure. Specifically, the lattice structures with slender geometry or made of less ductile materials possibly exhibit post-yield softening under compression [31–33]. Accordingly, to develop a robust lattice structure without post-yield softening, systematic studies on lattice structures made material with high plasticity are a pre-requisite [12].

Among various possible material choices, pure copper (Cu) is one of the desired candidates due to its high plasticity under mechanical loading. Combining its excellent thermal conductivity with the superior specific energy absorption of the lattice structures, this unique physical property synergy could potentially lead to singular applications in thermal management and heat exchanger scenarios. To date, most AM-built lattice structures are made by the laser-powder bed fusion (L-PBF) technique, largely due to the process's refined surface finish and design freedom. However, it is currently a challenge to fabricate pure Cu through the conventional L-PBF route, which typically employs an infra-red laser with a wavelength >1000 nm. This is because the Cu energy absorptivity drops significantly (<10 %) once the laser wavelength is above ~800 nm [34]. Together with this material's high thermal conductivity (~400 W/m-K), it is difficult to maintain a stable melt pool during fabrication. A preliminary study on the solid Cu parts built via L-PBF reported a high density of internal fusion defects agreeing with the previous hypotheses [35]. This problem can be partially alleviated by increasing the laser power [36,37], decreasing the spot size [38], and decreasing the powder size [39]. A more direct approach is to employ a laser source with a lower wavelength. It has been recently reported that the high energy absorptivity of Cu under a blue [40] or a green laser [41,42] ($\lambda$ = 450 and 515 nm, respectively) allows fully dense prints of Cu parts. Correspondingly, conceptual/theoretical studies for Cu-based 3-dimensional lattices, utilizing their mechanical, electrical, and thermal properties have been previously proposed [43–45], yet the fabrication and mechanical performance of Cu-based lattice structures remain unexplored so far.

To evaluate the mechanical performances and the energy absorption capacity of lattice structures, the deformation behavior needs to be investigated at a wide range of strain rates. This is because the strength of bulk metallic materials shows a distinctive strain rate dependency, governed by the dislocation movements [46]. Likewise, lattice structures made of such materials are also subjected to the influence of different strain ratesduring plastic deformations. However, due to the experimental complexities and the requirement of different testing devices, the deformation behavior of previously designed lattice structures was mainly studied either at quasi-static (0.001 /s) [47,48] or dynamic strain rates (1000 /s) [49,50]. Therefore, this is currently a knowledge gap on mechanical behavior of lattice structures across the full range of strain rates, i.e., including the quasi-static, the intermediate (0.01 ~ 100 /s), and the dynamic strain rates (>1000 /s). Such data is critical to thoroughly understand the deformation behavior of lattice structures and propose future design guidelines for lattice structure adoptions.

In this study, we systematically investigated the mechanical properties of pure Cu-based lattice structures fabricated successfully, for the first time, using L-PBF with a green laser beam source. Two different lattice structures showing different deformation mechanisms were chosen. The structure and microstructure of each lattice structure were examined using X-ray and electron microscopy-based methods, respectively. The lattice structures were compressed at a wide range of strain rates from 0.001 to 1000 /s. Combined with post-mortem structural and microstructural analysis, the deformation mechanisms and the mechanical performance of Cu lattice structures were identified. We believe that this study will expand the applicability of additively manufactured 3-dimensional architectures and will provide a guideline for the fabrication of near-ideal ductile metallic lattices for dynamic applications.

## 2. Experimental methods
### 2.1. Sample preparation

The deformation of strut-based lattice structures is primarily dictated by the nodal connectivity according to Maxwell's criteria [51],

$$M = b - 3j + 6 \tag{1}$$

where $b$ and $j$ are the number of struts and intersections (nodes) in the unit cell. For the structure with high nodal connectivity ($M \geq 0$), the deformation is dominated by the stretching of struts. For the structure with low nodal connectivity ($M < 0$), the deformation is dominated by the bending of struts [32]. Accordingly, we selected an octet-truss structure with $M = 0$ (denoted as Oct) and a cuboctahedron structure with $M = -6$ (denoted as Cub) as unit cells of the lattice structures (Figure 1(a, b)). The unit-cell length and the strut diameter were chosen as 1.5 mm and 300 μm, respectively. The relative densities of Oct and Cub structures are 0.38 and 0.21, respectively. Each lattice structure was designed to have 80-unit cell repetitions: A 4-repetition of unit cells along the z-axis, and a 20-repetition of unit cells in a cross-section perpendicular to the z-axis. In the cross-section perpendicular to the z-axis, we discretized the unit-cell geometry and arranged them symmetrically, optimizing the lattice geometry for the mechanical test (Figure S1). Additional plates at the top and the bottom of the lattice structure were introduced to ensure uniform deformation under loading along the z-axis. Computer-aided design (CAD) files of the Cub and the Oct structures were designed for the following fabrication process.

We utilized the L-PBF process with both green laser and infra-red laser beams (λ = 515 and 1064 nm) for the sample fabrication. The pure Cu (electrolytic tough-pitch) powder with a diameter of 16-63 μm (Nanoval GmbH) was used. Before the lattice fabrication, the L-PBF parameters, such as laser power, scan speed, layer thickness, and beam diameter were determined after a sequential process parameter optimization by fabricating the test cuboidal samples (10 mm × 10 mm × 10 mm) with varying parameters to maximize the density.

Afterwards, vector parameters such as beam compensation and gap between vector contours were optimized on the lattice structures. For the lattice fabrication with the green laser, laser power, scan speed, and beam diameter were set as 600 W, 1000 mm/s, and 160 μm, respectively. For the lattice fabrication with the infra-red laser, laser power, scan speed, and beam diameter were set as 600 W, 800 mm/s, and 80 μm, respectively. In both cases, a layer thickness was set as 20 μm. For the lattice part, due to the small strut diameter, only the contour laser scanning was adopted (contour fill). A gap of 80 μm was introduced between each contour. For the top and bottom plates, a hatch spacing of 100 μm was adopted. The LPBF process was conducted under inert gas (Ar) flow (< 100 ppm residual oxygen).

To reduce the surface roughness of as-fabricated lattice structures, a chemical etching step was conducted on every specimen before mechanical testing (Figure S2). The Cub and Oct copper structures were immersed into a chemical etchant (5 g $FeCl_3$ + 10 ml HCl + 100 ml $H_2O$) for 200 sec [52–54].

## 2.2. Rate-dependent compression tests

The mechanical properties of Cu lattice structures across six orders of strain rate magnitude were investigated. To explore such a wide range of strain rates, we utilized a combination of dilatometer (DIL 805A/D, TA instruments) (0.001/s – 10/s) and a Split-Hopkinson pressure bar (SHPB) testing setup (1000/s) at room temperature. The SHPB testing setup consists of two bars with a length and diameter of 1800 mm and 20 mm. Both the incident bar and the transmission bar are made of aluminum. The striker has a length of 500 mm and is made of the same material as the bars. With strain gages placed centrally on the bars, the signals can be recorded at a sampling rate of 1M samples per sec. Striker speeds between 8 and 14 m/s were used to achieve the required high strain rates and strains.

Before the compression test, the top and bottom surfaces of each specimen were mechanically polished. The structures were compressed to 25 and 50 % engineering strains at strain rates of 0.001, 0.01, 0.1, 1, 10, and 1000 /s. We evaluated the nominal stress and strain by dividing the load by the effective cross-sectional area of the lattice structure and by dividing the displacement by the height of the sample, respectively. The structural and microstructural analyses were conducted before and after the mechanical test to explore the deformation mechanisms involved.

## 2.3. Structural analysis

2D and 3D structural analysis of additively manufactured metal parts can be effectively studied using computed tomography (CT) [55,56]. ProCon CT-Alpha industrial system of X-ray CT system was used for the non-destructive structural analysis of the Cu lattice structures. This CT system includes a high-precision setup, leading to fine and stable geometric positioning within the 3D space using a 5-axis of x-y-z-rotation-tilting system with a reproducible accuracy < 1 µm and resulting in a resolution of up to around 2 µm for appropriate contrast images. The detector system XRD 1611 AP3 has 4064 × 4064 pixels, with each pixel dimension of ~100 µm. The X-ray tube XWT-240-TCHE Plus by X-ray WorkX GmbH has an anode made of tungsten target material. This X-ray tube reaches a maximum of 240 kV, leading to increased penetration of high-density samples. Software VG Studio MAX 3.5 of Volume Graphics GmbH (Heidelberg, Germany) was used for the 3D reconstruction and analysis of the MicroCT scans.

In this study, CT investigations were carried out using high-power X-ray tube mode to reach appropriate statistics within a short time. The typical parameter values for 3D image exposition were as follows: voltage of about 150 kV and current of about 200 µA during a measurement time of approximately 30-40 min. The voxel size for the samples investigated was around 30 µm and typical filters with zirconium plates of ~1mm thickness were used. These parameters were kept constant for all the investigated samples to ensure stable and comparable CT results.

### 2.4. Microstructural analysis

We characterized the microstructure of Cu lattice structures before and after the deformation. The cross-section of specimens parallel to the building direction was polished with colloidal silica. A field-emission scanning electron microscope (FE SEM, Sigma, ZEISS) equipped with an electron backscattering diffraction (EBSD, TSL) detector was used for obtaining the microstructure. A critical misorientation angle for grain boundaries was set as 15°, to distinguish between low-angle grain boundary (LAGB) and high-angle grain boundary (HAGB).

### 2.5. Finite element analysis

Stress distribution in the lattice structures under compression load was examined through finite element analysis (COMSOL Multiphysics). Based on the CAD design of Oct structure, a 3-dimensional model was constructed. Due to the high computational cost, only the elastic deformation of Oct structure was considered. Density, elastic modulus and Poisson's ratio of Cu were set as 8960 kg/m$^3$, 120 GPa and 0.34, respectively. To describe the deformation under compression, the bottom plate of the structure was fixed, and the top plate was set to move downward by the prescribed displacement. The Oct CAD model was subsequently meshed using triangular elements. There were 187,393 elements and 840,294 degrees of freedom in the model.

## 3. Results and Discussion

### 3.1 Structural analysis

The Cu lattice structures in this study show well-defined open-pore channels and nodal connectivity. As-fabricated lattice structures have rough surfaces originating from the unmelted powder adhesion, which may induce strong stress concentrations and facilitate subsequent crack initiation during deformation [57–59]. Therefore, chemical etching was conducted on the lattice structures. Figure S2 shows the reduction of unmelted powders on the surface after the chemical etching. Figure 1 (b) shows the optical photographs of undeformed Oct and Cub structures after the chemical etching. Visual inspection shows that the open-pore channels are clearly distinguished, and they are consistent with the pre-defined CAD design. Detailed geometries and the nodal connectivity of lattice structures can be obtained from the tomographic analysis (Figure 2). Figure 2 (a, b) are the 3-dimensional reconstruction slice of each structure viewed from a specific angle where the Oct and Cub structures can be easily distinguished (detailed CT 3-dimensional reconstructions of lattice structures are shown in Movies M1 and M2). Importantly, the horizontal and the inclined struts are almost identical to the original CAD designs that constitute and differentiate the lattice structures from each other. The geometrical feature sizes of each structure were also evaluated from the reconstructions. From the 3-dimensional reconstruction, the unit-cell length and the strut diameter of Oct and Cub structures are 1.46 (± 0.02) and 1.49 (± 0.01) mm, 328 (± 27) and 347 (± 16) μm, respectively. Compared to the original designs (1.5 mm of unit-cell length and 300 μm of strut diameter), the printed parts show only 2 and 13 % difference in the unit-cell length and the strut diameter of Oct and Cub structures.

Both structures show rough surfaces with spherical agglomerates as confirmed by the surface SEM images (Figure S2). The spherical agglomerates may stem from a balling effect induced by less optimized process parameters [60,61]. Moreover, the fabrication of 3-dimensional structures with open-pore channels through L-PBF leads to uncontrolled and inhomogeneous powder adhesion on the beam surfaces and resultant rough surfaces [62–64]. The high thermal conductivity of Cu leads to further unmelted powder adhesion to the melt pool during solidification due to insufficient remelting [65,66].

The presence of internal pores in both structures was confirmed via experimental X-ray CT investigation. Figure 2 (c, d) shows a CT analysis of the porosity distribution in lattice structures. The CT study provided detailed 3-dimensional pore distributions in the lattice structures, as shown in Movies M1 and M2. In both cases, the volume of closed pores is mostly below 0.01 mm$^3$, but there are also a few large pores with a volume of 0.03 ~ 0.05 mm$^3$. The closed pores with sizes below 0.01 mm$^3$ are most probably due to the lack of powder fusion [67,68]. Notably, there are more small pores in the Cub structure than in the Oct structure. It is known that structures with low inclination angles fabricated by the L-PBF process tend to exhibit high porosity [69]. It is attributed to a different cooling rate in the upper part and lower part of the strut with a low inclination angle. The excess energy in the lower part may lead to an instability of the melt pool. The Cub structure shows a higher volume fraction of horizontal struts (33 %) compared to the Oct structure (25 %). Furthermore, the Oct structure shows higher nodal connectivity which may lead to a relatively uniform cooling rate in the horizontal struts. Of the large pores in both structures, most of them are located in the lower half of the structure. Given that the bottom plate of the lattice structure was in direct contact with a base substrate, the cooling rate at the bottom plate may be slightly higher than that at the lattice structure and the top plate. Such a higher cooling rate may reduce the heat absorption of material during fabrication, leading to the formation of pores.

Even with the surface roughness and the internal pores, the lattice structures in this study which are fabricated by green laser irradiation show considerably fewer fusion defects compared to the previously studied pure Cu structures which were fabricated by using a conventional L-PBF process with infra-red laser [37,70–76]. As a direct comparative study, we also fabricated the Oct structure by using the conventional L-PBF process with an infra-red laser (Figure S3a,b). Due to the low energy absorption, it exhibits even larger fusion defects, and the defects/cracks are distributed throughout the structure (Figure S3) [35]. To summarize, the high energy absorption of Cu from the green laser during the L-PBF process is the key to obtaining a well-defined geometry and a low defect density in lattice structures.

## 3.2 Microstructural analysis

The materials fabricated by L-PBF typically show unique microstructures such as columnar dendritic structures [77–79], elongated grains [80–82], strong texture [83–85], and dislocation cell structures [86–88] along the building direction due to the repetitive rapid heating and cooling. Hence, we examined the microstructure in the cross-section along the building direction of each lattice structure.

Figure 3 shows the microstructure of the representative unit cross-section in the undeformed Oct structure. Consistent with the structural analysis using microCT, the cross-section shows the well-defined and fully dense struts and nodes. From the inverse pole figure (IPF) maps along the building direction and the corresponding pole figure maps in Figure 3 (a), the microstructural characteristics of lattice structures were identified. Firstly, there is no dendritic structure in the material. The columnar dendrite formation originates from the slow growth process during the L-PBF process [89,90]. We hypothesize that the large thermal gradient and the rapid solidification rate of Cu prevent the formation of columnar dendrite [91,92]. Secondly, there is no strong crystallographic texture in the material. The pole figure maps along the building direction in Figure 3 (a) indicate that there is no apparent preferential texture. Metallic materials fabricated by the L-PBF process are known to show a strong crystallographic texture along the building direction, as the grain grows parallel to the temperature gradient [83–85]. For bulk copper, {100} or {110} texture along the building direction was reported [93]. However, for the lattice structures, the unmelted or partially melted particles act as nucleation sites for new grains, which hinders the texture development along the building direction [94]. Also, multiple re-melting and solidification cycles can introduce randomness in the crystallographic texture [93]. The small difference between the gap of each contour scan (80 μm) and the laser beam diameter (160 μm) used for the lattice fabrication process leads to additional re-melting and solidification. Thirdly, we hypothesize that there are residual stresses at the node of the lattice structure. The grain boundary map and Kernel average map in Figure 3 (b) indicate residual stress of solidified materials [95,96]. At the nodes of the lattice structure, there is a high density of low-angle grain boundary and high local misorientation. Thermal stress at the node may not have been relieved as the volume change is limited by the surrounding truss components. The misorientation angle distribution in Figure 3 (c) confirms a high fraction of LAGB, while for HAGB (> 15°), the distribution follows that of random misorientation. Lastly, the average grain size (measured by the equivalent circle diameter method) of the lattice structure is 30 ± 19 μm. The grain size distribution in Figure 3 (d) implies a large fraction of small size grains (< 15 μm). As shown in Figure 3 (a, b), the small grains are mostly located at the outer surface of the structure (thickness of small grain layer is 38 ± 10 µm). These grains may come from unmelted or partially melted powders remaining after the chemical etching. The chemical etching was conducted on the as-fabricated lattice structures to partially alleviate the surface roughness which may lead to early yielding or crack formation during deformation. Nevertheless, some unmelted powder remains even after chemical etching. Though, it is expected that these partially attached powders with small grains on the surface are unlikely to significantly contribute to the strength of the lattice structures. We also investigated the microstructure of the Cub structure as shown in Figure S4. Similarly, the Cub structure shows elongated grains without strong texture, LAGB at the node, and average grain size of 30 ± 20 µm. As such, it could be concluded that a nominal change in the design/geometry does not significantly affect the microstructure of the Cu lattices.

**3.3 Deformation behavior**

To investigate the applicability of the Cu lattice structure in dynamic applications, we conducted compression tests at various strain rates. Figure 4 (a, b) shows the engineering stress-strain curves of Oct and Cub structures, respectively, at 0.001, 0.01, 0.1, 1, 10, and 1000 /s. Several distinct deformation characteristics can be observed in the stress-strain curves. Firstly, at every strain rate, the Oct structure shows about a 1.5-time higher stress level than the Cub structure. The difference in stress level stems from the deformation mechanism of each structure. As mentioned above, the deformation of the lattice structure is governed by the connectivity of struts. Hence, the Oct structure with higher connectivity is stiffer and stronger than the Cub structure with lower connectivity. Furthermore, the higher relative density of the Oct structure than that of the Cub structure results in a higher stress level. Secondly, there is no post-yield softening in the stress-strain curves of the Oct structure. Previous studies on the Oct structure made of Ti-6Al-4V and polymer reported post-yield softening during deformation [31–33]. This deformation behavior originates from the high connectivity of the structure which leads to an abrupt buckling or fracture of struts. The deformation process also depends on the exact dimensions and the material choice. The buckling of the strut is known to occur in slender struts [97,98]. Tancogne-Dejean *et al*. computationally demonstrated that the Oct structures with low relative density, i.e., high strut aspect ratio, are prone to buckling under compressive stress [98]. They identified that there was no buckling deformation in the Oct structures with strut aspect ratio lower than 5. For the lattice structures in this study, the aspect ratio is 3.54, hence the struts are not favorable for buckling deformation. In addition, Cu which typically possesses high ductility may enable a gradual plastic deformation in the struts. Therefore, the post-yield softening of the Cu Oct structure is less probable despite the high strut connectivity. As hypothesized, we observed a continuous and smooth hardening behavior in both Oct and Cub Cu structures during deformation and it can be attributed to the intrinsic mechanical properties of Cu. It has been reported that coarse-grained Cu with a grain size of 10-50 μm shows a high strain hardening exponent and high hardening capability [99,100]. Figure S5 shows tensile stress-strain curves obtained from dog bone shaped specimens oriented along the building direction of L-PBF. The L-PBFed Cu in this study shows a higher hardening exponent (n=0.27 ± 0.01) and failure strain (46 ± 3 %) compared to Ti- [27,28] and Al- [29] alloys previously used for the lattice fabrication. This intrinsic material property could also be a contributing factor toward the continuous and smooth hardening-based deformation of the Cu lattices. Third, the lattice structures show a clear strain rate dependency under compression. Both the structures show similar yield and flow stress levels even when the strain rate is increased from 0.001 /s to 1 /s. At 10 /s though, the flow stress slightly increases, and at 1000/s, both the yield and flow stress drastically increase. For the quantitative analysis of the lattice performance, we evaluated the 2% offset yield stress ($\sigma_y$), the flow stress at 20 % strain ($\sigma_{0.2}$), and the energy absorption capacity ($U_A$) of each structure compressed at various strain rates. Figure 4 (c, d) shows the comparison of $\sigma_y$ and $\sigma_{0.2}$. The strain rate sensitivity (*m*) was calculated based on the flow stress as below:

$$m = \frac{\partial \ln \sigma_f}{\partial \ln \dot{\varepsilon}} \tag{2}$$

where $\sigma_f$ is the flow stress and $\dot{\varepsilon}$ is the strain rate. In order to evaluate *m* in the plastic deformation regime, the flow stresses at 0.2 strain were selected. It was reported that the cellular materials with relative density higher than 0.4 commonly exhibits *m* close to based materials. On the other hand, it was reported that the cellular materials with relative density lower than 0.4 shows increased *m* due to the micro-inertia effect [101–103]. For the Oct and Cub lattices, *m* were calculated as 0.016 and 0.028 respectively. The *m* of Oct structure (relative density of 0.38) is analogous to the polycrystalline bulk Cu (m = 0.0158) [104,105]. The Cub lattice with relative density of 0.21 shows increased *m* compared to the Oct lattice. The sharp increase in $\sigma_y$ and $\sigma_{0.2}$ at 1000 /s can be attributed to a transition in the deformation mechanisms. It was reported previously that the polycrystalline Cu shows distinctive strain rate sensitivity in compressive strength, especially at high strain rates [104,105]. At dynamic strain rates (> 100 /s), the dislocation motion is affected by the phonon vibration and the viscous-drag effect of dislocation by phonons during the plastic deformation leads to a sharp increase in the $\sigma_y$ and $\sigma_f$ [106]. Consequently, there is a clear increase in $\sigma_y$ and $\sigma_{0.2}$ when the strain rate is increased from 10 /s to 1000 /s, suggesting a possible viscous-drag effect in the Cu lattice structures. The $U_A$, as shown in Figure 4 (e), can be obtained as follows:

$$U_A = \int_0^{\varepsilon_d} \sigma \, d\varepsilon \tag{3}$$

where the $\sigma$, $\varepsilon$, and $\varepsilon_d$ are the stress, strain, and densification strain. The $\varepsilon_d$ is the critical strain where the open pore channels are closed, and at $\varepsilon > \varepsilon_d$ the flow stress drastically increases. The lattice structures in this study show continuous hardening during compression, obscuring trivial identification of $\varepsilon_d$. From the microCT scan image of the Cub structure, it can be seen that densification occurred at samples compressed up to 50% strain (Figure S6). To quantitatively determine $\varepsilon_d$ from the stress-strain curves, the second derivative was taken to obtain the concavity of the curve. The strain where the second derivative becomes positive and drastically increases is selected as $\varepsilon_d$ (Figure S7). The densification strain of Oct and Cub structures was identified as 41 % and 45 %, respectively. The difference in densification strain stems from the different relative densities of structures. Figure 4 (e) clearly shows that $U_A$ of Oct structure is significantly higher than that of Cub structure due to the higher stress level, originating from differences in the nodal connectivity and the relative density. Furthermore, there are sharp increases in $U_A$ of Oct and Cub structure at 1000 /s strain rate, which can be attributed to the viscous drag effect.

We further investigated the structural evolution in the deformed lattices. Figure 5 shows the MicroCT scan images of undeformed and compressed Oct and Cub structures at 0.001, 1, and 1000 /s strain rates at 25% strain. The images were obtained at the central cross-section of structures. The scan images of undeformed Oct and Cub structures (Figure 5 (a, b)) show well-defined open-pore channels and struts. Notably, there are also low contrast areas at the bottom layers of each lattice structure (denoted by yellow arrows), which most probably originate from the residual unmelted powders [107]. The open-pore channels also can be identified in the scan images of compressed structures at 0.001 /s (Figure 5 (c, d)), meaning that the structures are not densified yet at a strain of 25%. This remains the same for other compressed structures at higher strain rates of 1/s (Figure 5 (e, f)) and 1000 /s (Figure 5 (g)). Moreover, the scan images of compressed Oct and Cub structures commonly show that the structures were compressed homogeneously, suggesting that there was no localized deformation. In previous studies on the lattice structures, a localized shear band formation in the structure during compression was reported [108–110]. The shear bands crossing the entire structure are typically formed with a crack initiation or severe deformation of the strut, decreasing the energy absorption capacity. This localized deformation can be suppressed by improving the ductility of materials of lattice structures [110,111]. Hence, it can be deduced that the ductile Cu enables homogeneous deformation, and the homogeneous deformation even at dynamic strain rates results in robust energy absorption by the Cu lattice structures.

We also investigated the possible microstructural evolution in the deformed lattice structures. Especially because the deformation at dynamic strain rates is known to introduce unique microstructures such as a strong shear localization and dynamic recrystallization to materials [112,113]. To investigate the microstructural change, we examined the cross-section along the building direction of Oct structures compressed at quasi-static and dynamic conditions and compared the microstructures (Figure 6). The post-mortem microstructure analysis confirms that there is no severe inhomogeneous deformation induced during dynamic compression. In the grain boundary map of deformed materials, the LAGB indicates plastic deformation [95]. Figure 6 (a) shows the grain boundary map overlayed with the image quality map of the representative unit cross-section of the Oct structure compressed at 0.001 /s and 1000 /s, respectively. Notably, there is no clear difference in the grain boundary maps, suggesting that dynamic compression may not induce inhomogeneous deformation. Instead, compared to the microstructure of the undeformed Oct structure in Figure 3(b), regardless of the strain rate, the compressed structures show increased LAGB at the points where the strut and the node are connected. This LAGB may originate from the stress concentration and resulting deformation under compressive load. As confirmed by the finite element analysis (Figure S8), when the structure is compressed, the stress is concentrated at these connecting points. Consistently, the corresponding KAM maps (Figure 6 (b)) show high local misorientation at the strut-node connecting area. Figure 6 (d) shows that there is no apparent change in misorientation angle distribution with strain rate. Compared to the distribution of undeformed Oct structure in Figure 3(b), the compressed structures again show an increase in the low-angle grain boundaries (< 15°) which can be attributed to the plastic deformation[95].

We hypothesize that dynamic recrystallization was unlikely to occur in these Cu architectures during dynamic compression. Figure 6 (c) shows grain orientation spread (GOS) maps overlayed with the image quality map of the Oct structure compressed at 0.001 /s and 1000 /s. To separate the recrystallized grains from the deformed grains, a GOS cutoff of 1 deg was taken. There is no clear difference between the Oct structures compressed at 0.001 /s and 1000 /s. Moreover, there is no apparent indication that the recrystallization occurred at the strut-node connecting area where the stress is concentrated. The grain size distribution in Figure 6 (e) shows no difference between structures compressed at 0.001/s and 1000/s. It is also supported by no softening in the stress-strain curves (Figure 4 (a, b)), a typical signature of dynamic recrystallization. It was reported that the adiabatic temperature rise during deformation at a high strain rate induces dynamic recrystallization of bulk Cu at a dynamic strain rate of $10^4$ /s [114,115]. With a relatively low dynamic strain rate ($10^3$ /s) and pore channels in the lattice structure made of thermally conductive Cu, the adiabatic temperature rise required for the dynamic recrystallization may not be favorable.

The specific compressive strength and energy absorption capacity of Cu Oct and Cub structures in this study were compared to other structures made of Cu [116–118], Ti alloy [27,28], Al alloy [29], stainless steel [30], and alumina [24–26] from previous reports and are summarized in Figure 7. As shown in Figure 7 (a), the compressive strengths of the Oct and Cub structures at 0.001 /s in this study are in an analogous trend to the previous studies on the pure Cu micro-and nano-scale structures [116–118]. Notably, in these small structures, the material size effect may play a significant role in the mechanical properties. However, in the case of the nano-foams, the absence of periodicity reduces the load-bearing capability of the structure [117,118]. In the case of the micro Oct structure, due to the grain size which is comparable to the strut diameter, it shows a single crystalline-like deformation behavior [116]. Hence, even though the macroscale lattice structures of this study are 2-3 orders of magnitude larger than the microlattices tested in previous studies, they exhibit comparable strength due to the combination of polycrystalline microstructure and ordered architecture.

Further, the alumina hollow Oct structures show extremely high specific strength (10 times higher than the Cu lattice structures) due to the low density and the high compressive strength (Figure 7 (a)) [24–26]. However, due to its intrinsic brittleness, the energy absorption capacity is as low as the metal foams as shown in Figure 7 (b). On the other hand, the Cu Oct and Cub structures, due to their ductile nature, show high energy absorption capacity, outperforming the strong alumina structures. Moreover, as the Cu lattice structures show continuous strain hardening without post-yield softening, the energy absorption capacity is comparable to that of lattice structures made of high-strength alloys. As such, the current study on the lattice structures made of Cu and their mechanical properties at a wide range of strain rates provides evidence for the applicability of ductile metals as appropriate structural materials for dynamic energy absorption. In the future, the surface roughness of lattice structures, which may lower their mechanical performance, can be further improved by parameter optimization via high-throughput screening [119], by adopting a multi-beam exposure strategy for lattice fabrication [120,121] or by conducting additional post-treatment such as laser polishing [122]. In summary, this study shows the feasibility of Cu 3-dimensional architectures in future multifunctional applications that require excellent mechanical, electrical, and thermal properties simultaneously.

## 4. Conclusion

We fabricated pure Cu lattice structures with different nodal connectivities using a green laser beam LPBF process. The structural and microstructural analysis demonstrate that the Cu lattice structures possess well-defined open-pore channels with low defect density. The mechanical properties under compression were investigated at a wide range of strain rates (0.001 ~ 1000 /s). The appropriate geometrical dimensions of the lattices prevented buckling deformation of the strut and the copper with its high ductility enabled continuous plastic deformation with strain hardening and prevented post-yield softening. Therefore, a high energy absorption capacity was identified, demonstrating the applicability of Cu lattice architectures as structural materials that can sustain dynamic loadings.

## Acknowledgements

SGK gratefully acknowledges funding by the National Research Foundation of Korea (NRF, No. NRF-2020R1A6A3A03039038). The authors would like to thank Michael Adamek for his assistance with the mechanical test. ZS acknowledges financial support from the Individual Research Grant (Grant reference No.:A20E7c0109) of the Agency for Science, Technology and Research of Singapore.

[30] B.K. Lee, K.J. Kang, A parametric study on compressive characteristics of Wire-woven bulk Kagome truss cores, Compos Struct. 92 (2010) 445–453. https://doi.org/10.1016/J.COMPSTRUCT.2009.08.029.

[31] L. Dong, V. Deshpande, H. Wadley, Mechanical response of Ti–6Al–4V octet-truss lattice structures, Int J Solids Struct. 60–61 (2015) 107–124. https://doi.org/10.1016/J.IJSOLSTR.2015.02.020.

[32] M.F. Ashby, The properties of foams and lattices, Philosophical Transactions of The Royal Society A. 364 (2006) 15–30. https://doi.org/10.1098/rsta.2005.1678.

[33] R. Schwaiger, L.R. Meza, X. Li, The extreme mechanics of micro- and nanoarchitected materials, MRS Bull. 44 (2019) 758–765. https://doi.org/10.1557/mrs.2019.230.

[34] D. Franco, J.P. Oliveira, T.G. Santos, R.M. Miranda, Analysis of copper sheets welded by fiber laser with beam oscillation, Opt Laser Technol. 133 (2021) 106563. https://doi.org/10.1016/J.OPTLASTEC.2020.106563.

[35] Q. Jiang, P. Zhang, Z. Yu, H. Shi, D. Wu, H. Yan, X. Ye, Q. Lu, Y. Tian, A review on additive manufacturing of pure copper, Coatings. 11 (2021). https://doi.org/10.3390/coatings11060740.

[36] M. Colopi, A. Gökhan Demir, L. Caprio, B. Previtali, Limits and solutions in processing pure Cu via selective laser melting using a high-power single-mode fiber laser, The International Journal of Advanced Manufacturing Technology. 104 (2019) 2473–2486. https://doi.org/10.1007/s00170-019-04015-3.

[37] T.-T. Ikeshoji, K. Nakamura, M. Yonehara, K. Imai, H. Kyogoku, Selective Laser Melting of Pure Copper, JOM. 70 (2018) 103–130. https://doi.org/10.1007/s11837-017-2695-x.

[38] S. Qu, J. Ding, J. Fu, M. Fu, B. Zhang, X. Song, High-precision laser powder bed fusion processing of pure copper, Addit Manuf. 48 (2021) 102417. https://doi.org/10.1016/J.ADDMA.2021.102417.

[39] M. Bonesso, P. Rebesan, C. Gennari, S. Mancin, R. Dima, A. Pepato, I. Calliari, Effect of Particle Size Distribution on Laser Powder Bed Fusion Manufacturability of Copper, BHM Berg- Und Hüttenmännische Monatshefte. 166 (2021) 256–262. https://doi.org/10.1007/s00501-021-01107-0.

[40] W. Hongze, K. Yosuke, Y. Ryohei, N. Yuya, S. Kunio, Development of a high-power blue laser (445 nm) for material processing, Opt Lett. 42 (2017) 2251–2254.
21

# Figures

**Figure 1.** Geometry of lattice structures made of pure Cu.(a) Computer-aided design of unit-cells and lattice structures. (b) Optical photographs of pure Cu lattice structures fabricated by L-PBF with green laser beam.

**Figure 2.** Tomographic analysis of pure Cu lattice structures. (a, b) 3-dimensional reconstruction of lattice structures. (c, d) Porosity distribution map of lattice structures.

**Figure 3.** Microstructure of representative unit cross-section in Oct structure. (a) Inverse pole figure and corresponding pole figure maps along building direction. (b) Grain boundary map and Kernal average misorientation map. (c) Misorientation angle distribution. (d) Grain size distribution plots.

**Figure 4.** Mechanical properties of pure Cu lattice structures. (a, b) Nominal compressive stress-strain curves of Oct and Cub structures at wide range of strain rates. Comparison of (c) 2 % offset yield stress, (d) 20 % flow stress, and (e) absorbed energy of Oct and Cub structures under compressive deformation.

**Figure 5.** Cross-sectional microCT images of Oct (left) and Cub (right) structures. (a, b) MicroCT images before compression. Yellow arrows indicate the unmolten powders which have negligible effect on the mechanical properties. (c-g) MicroCT images after compression to 25 % engineering strain at strain rates of 0.001, 1, and 1000 /s. (h) Schematics of cross-section of interest in each structure.

**Figure 6.** Microstructure of representative unit cross-section in Oct structure after compression. (a, b) Grain boundary map. (c) Misorientation angle distribution and (d) grain size distribution plots.

**Figure 7.** Comparison of mechanical properties of pure Cu lattice structures with previously reported values. (a) Compressive strength with respect to density. (b) Absorbed energy with respect to density.

**Figure 1.** Geometry of lattice structures made of pure Cu. (a) Computer-aided design of unit-cells and lattice structures. (b) Optical photographs of pure Cu lattice structures fabricated by L-PBF with green laser beam.

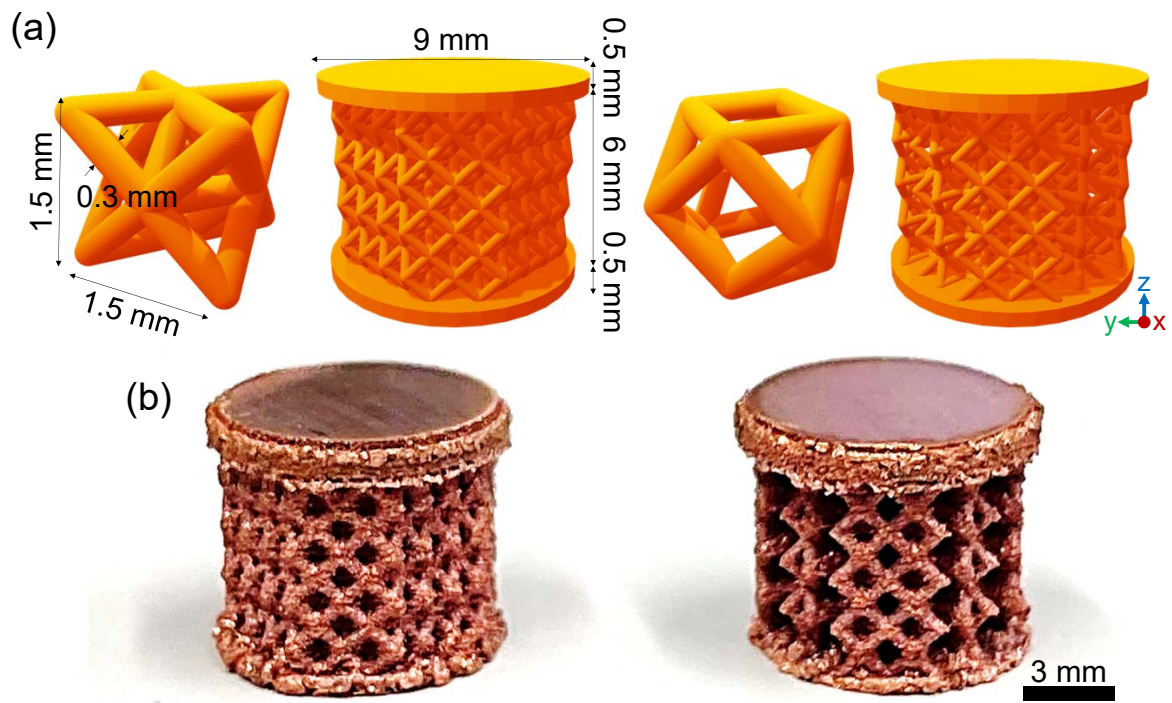

**Figure 2.** Tomographic analysis of pure Cu lattice structures. (a, b) 3-dimensional reconstruction of lattice structures. (c, d) Porosity distribution map of lattice structures.

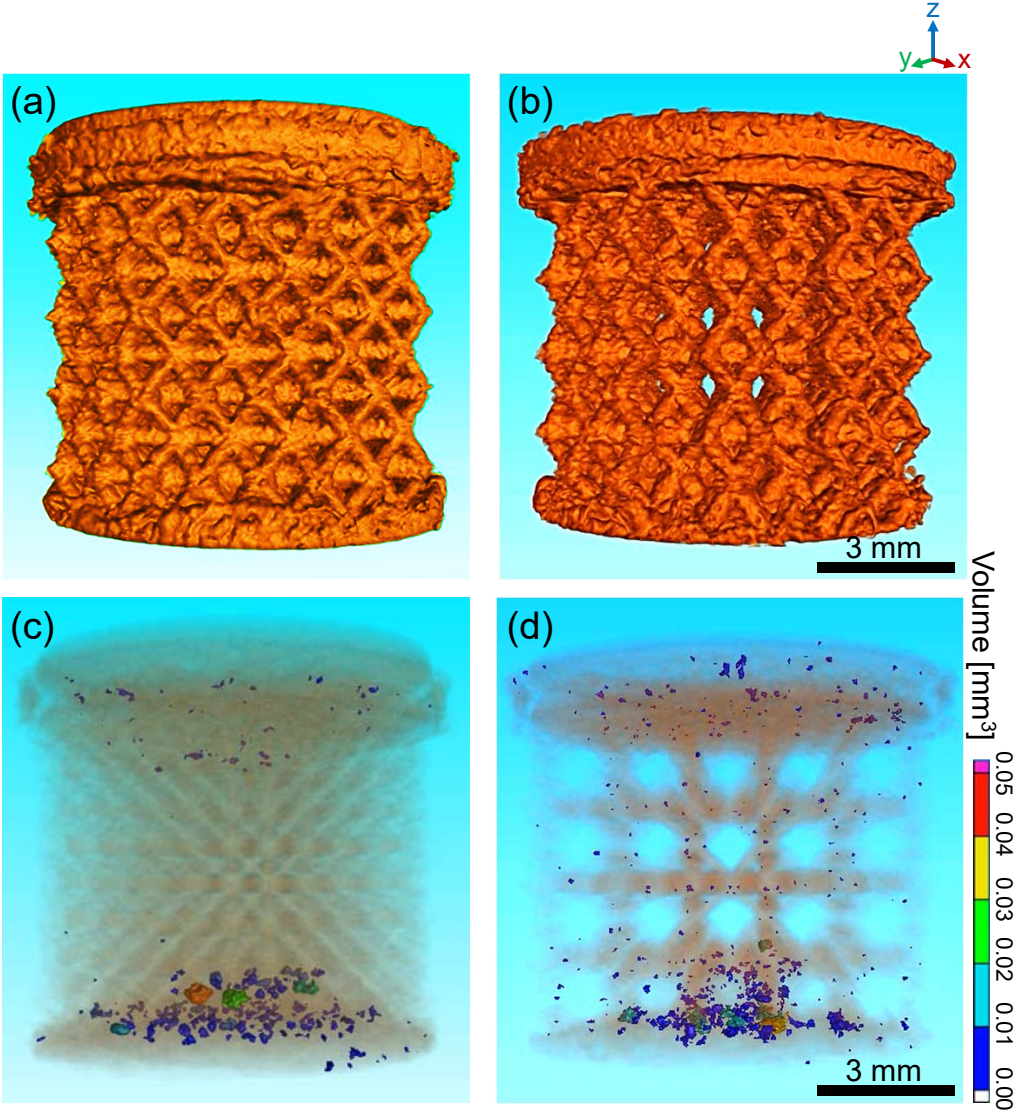

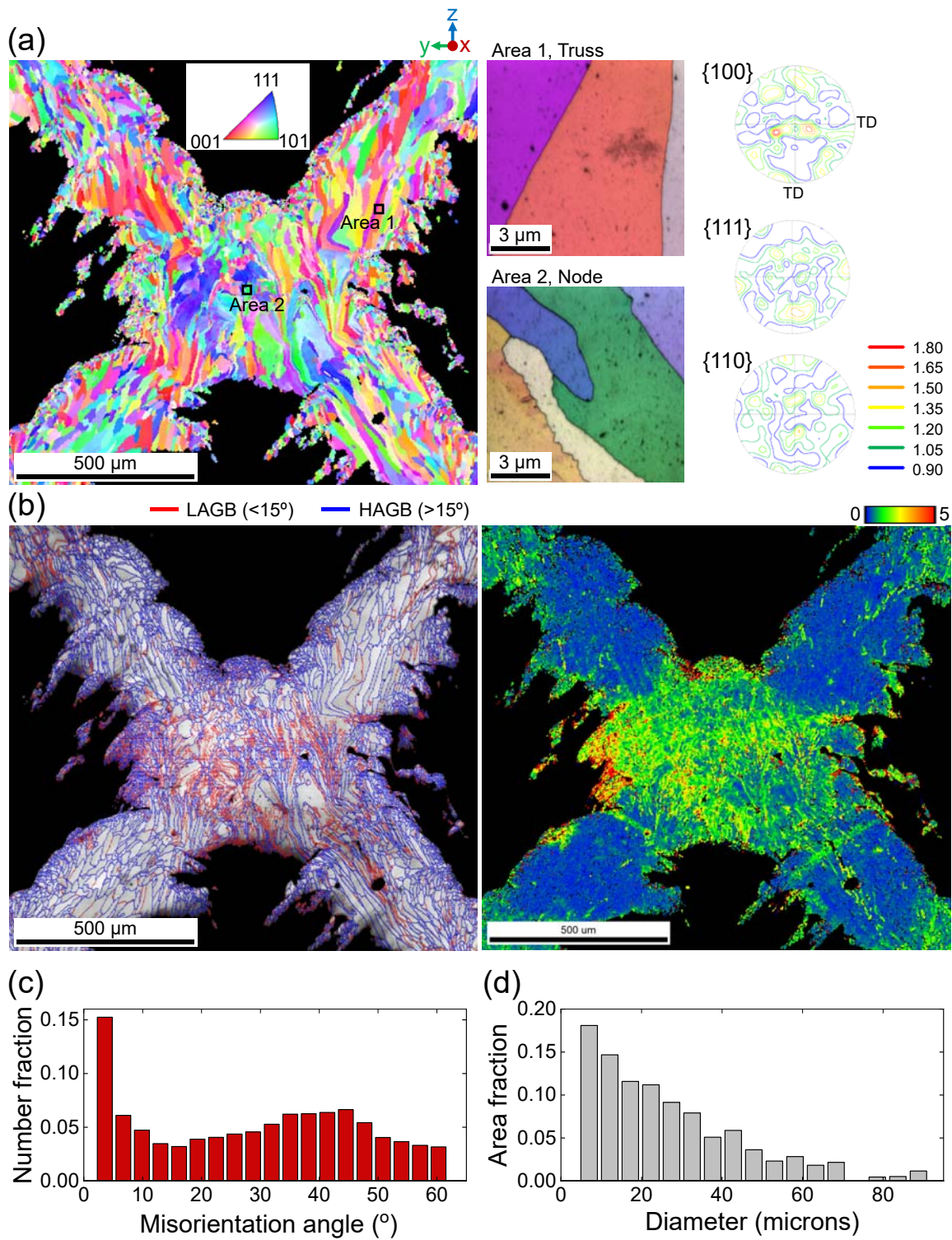

**Figure 3.** Microstructure of representative unit cross-section in Oct structure. (a) Inverse pole figure and corresponding pole figure maps along building direction. (b) Grain boundary map and Kernal average misorientation map. (c) Misorientation angle distribution. (d) Grain size distribution plots.

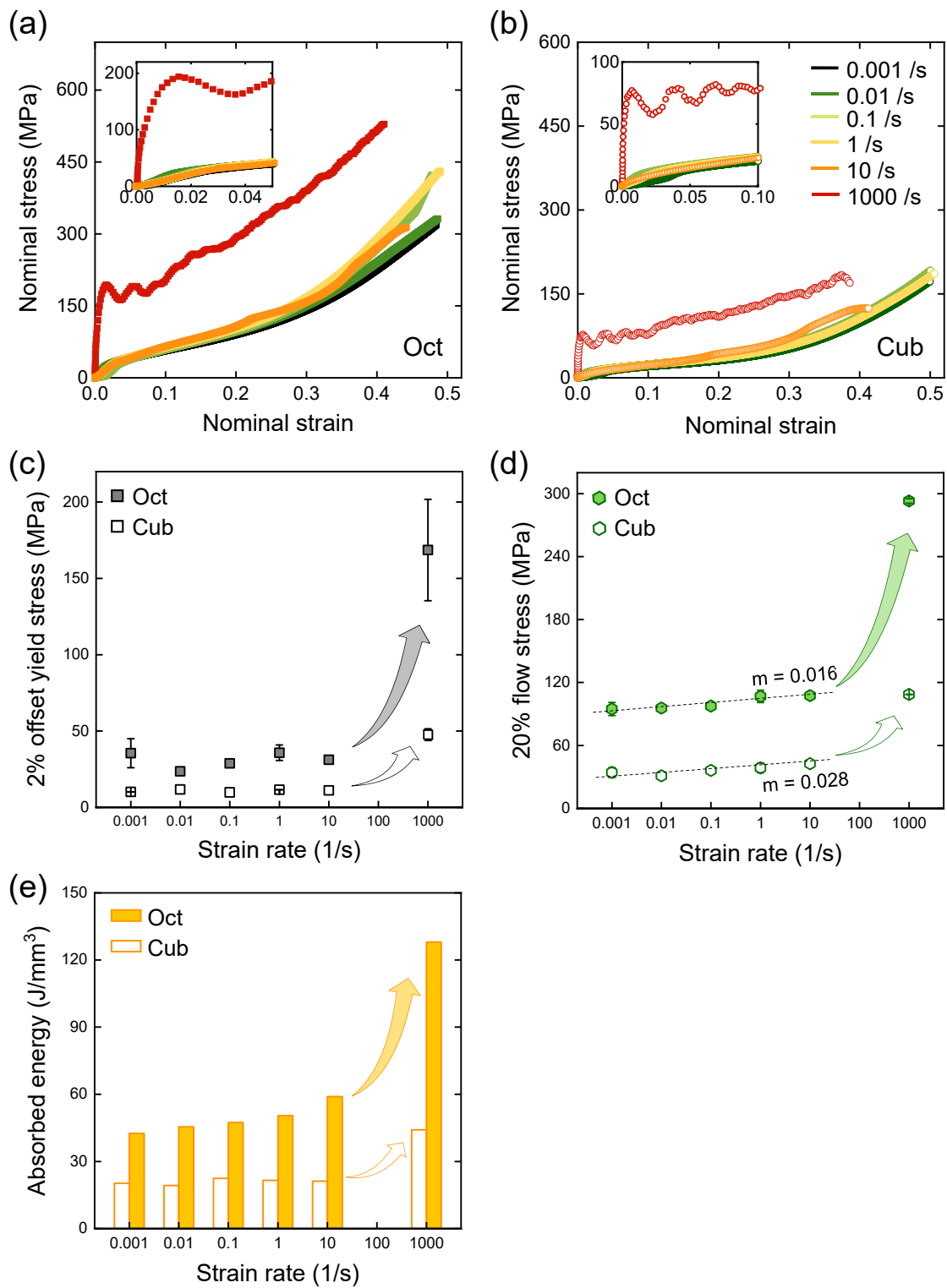

**Figure 4.** Mechanical properties of pure Cu lattice structures. (a, b) Nominal compressive stress-strain curves of Oct and Cub structures at wide range of strain rates. Comparison of (c) 2 % offset yield stress, (d) 20 % flow stress, and (e) absorbed energy of Oct and Cub structures under compressive deformation.

**Figure 5.** Cross-sectional microCT images of Oct (left) and Cub (right) structures. (a, b) MicroCT images before compression. Yellow arrows indicate the unmolten powders which have negligible effect on the mechanical properties. (c-g) MicroCT images after compression to 25 % engineering strain at strain rates of 0.001, 1, and 1000 /s. (h) Schematics of cross-section of interest in each structure.

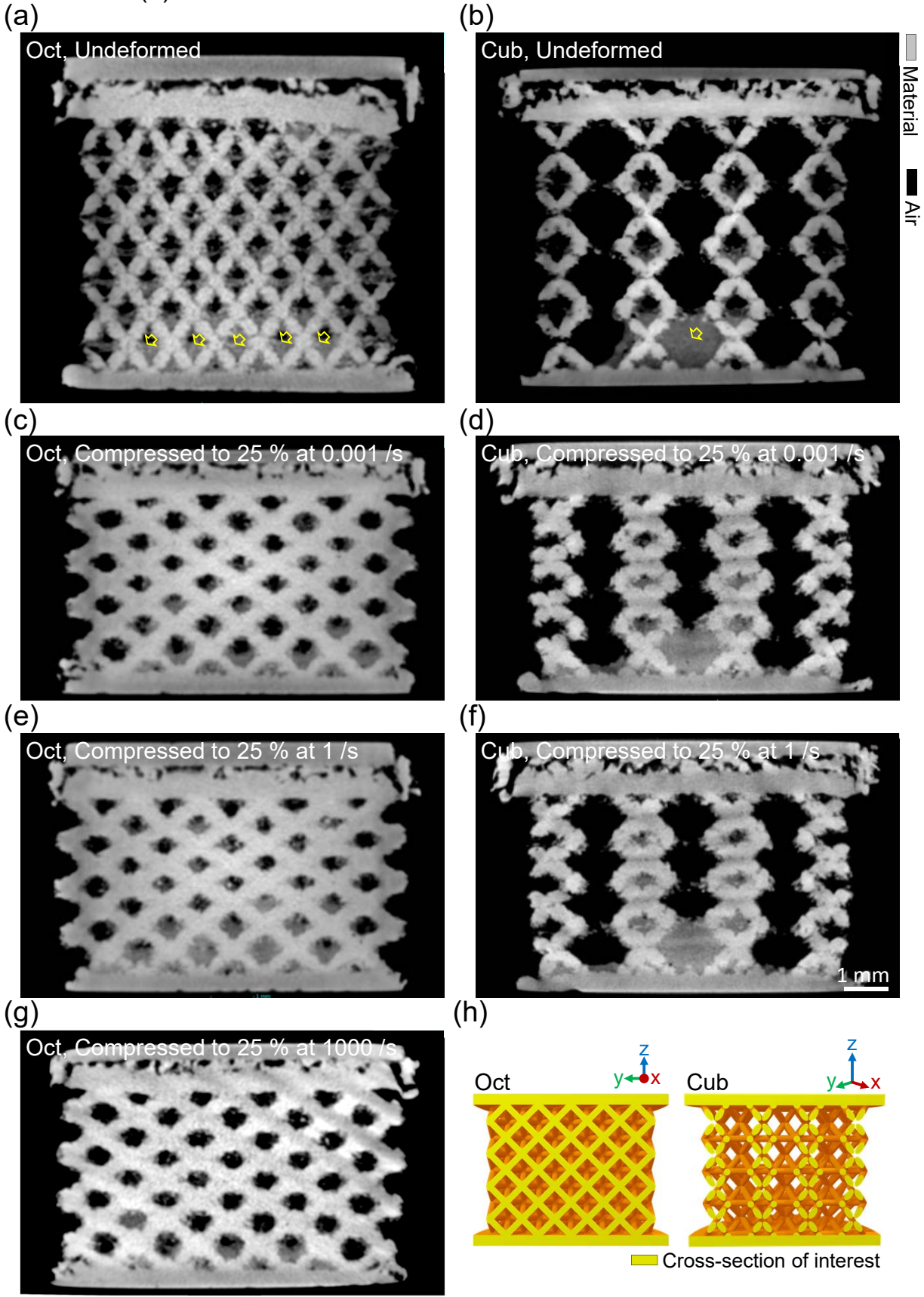

**Figure 6.** Microstructure of representative unit cross-section in Oct structure after compression. (a, b) Grain boundary map. (c) Misorientation angle distribution and (d) grain size distribution plots.

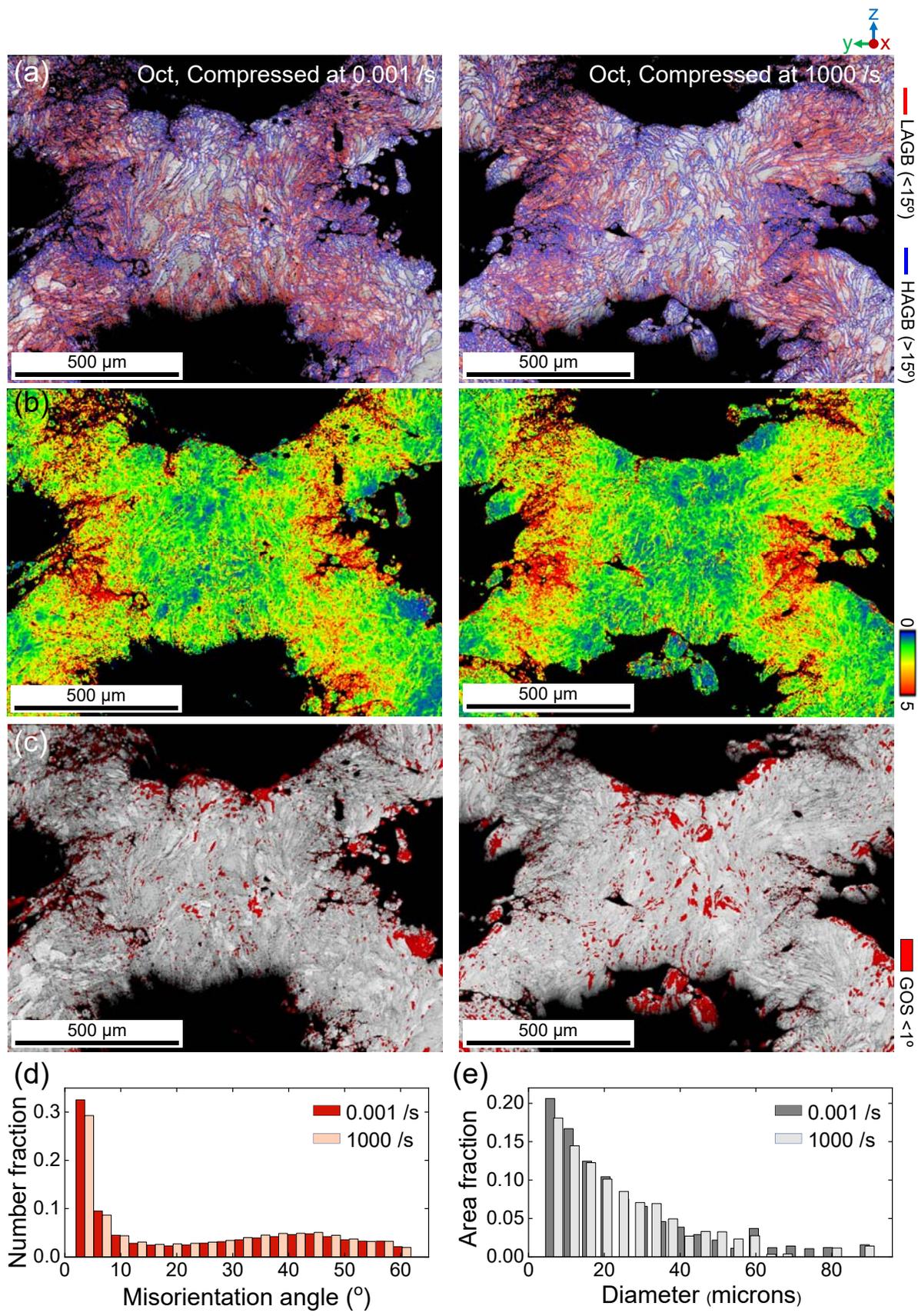

**Figure 7.** Comparison of mechanical properties of pure Cu lattice structures with previously reported values. (a) Compressive strength with respect to density. (b) Absorbed energy with respect to density.

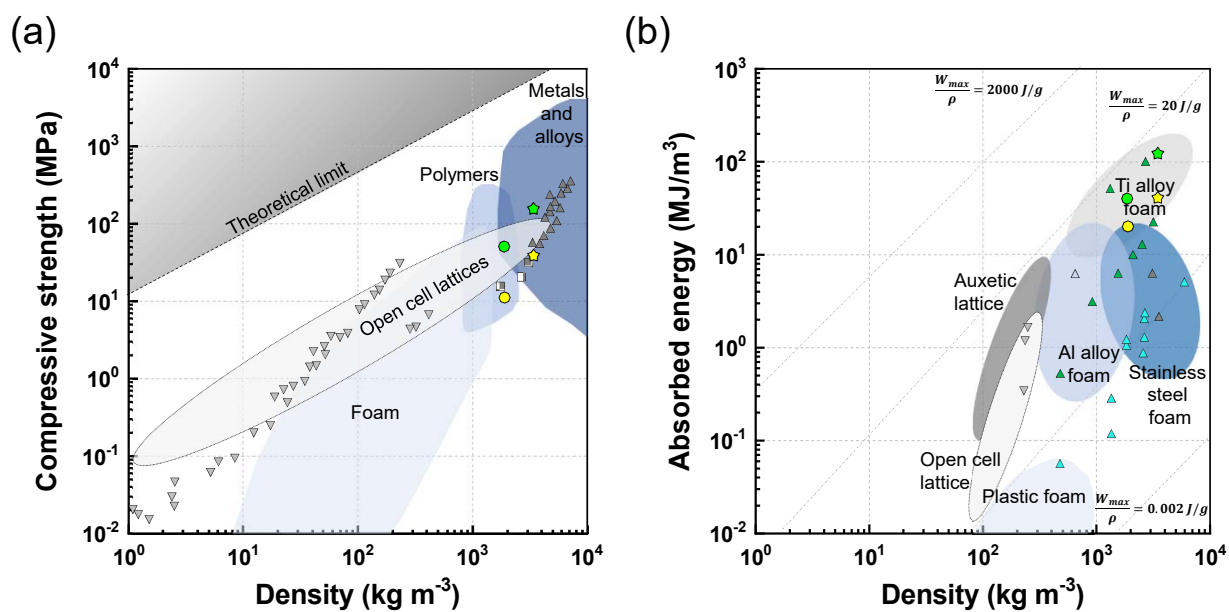

# Supplementary Information

Green laser powder bed fusion based fabrication and rate-dependent mechanical properties of copper lattices

Sung-Gyu Kang [a*], Ramil Gainov [b], Daniel Heußen [c], Sören Bieler [d], Zhongji Sun [e], Kerstin Weinberg [d], Gerhard Dehm [a], and Rajaprakash Ramachandramoorthy [a*]

[a] Max-Planck-Institut für Eisenforschung GmbH, Max-Planck-Straße 1, 40237 Düsseldorf, Germany

[b] RWTH Aachen University, Wüllner-Straße 2, 52062 Aachen, Germany

[c] Fraunhofer-Institut für Lasertechnik ILT, Steinbach-Straße. 15, 52074 Aachen, Germany

[d] Lehrstuhl für Festkörpermechanik, Universität Siegen, Paul-Bonatz-Straße 9-11, 57068 Siegen, Germany

[e] Institute of Materials Research and Engineering, Fusionopolis Way 2, 138634 Singapore

*Corresponding author: Dr. Sung-Gyu Kang and Dr. Rajaprakash Ramachandramoorthy

E-mail: s.kang@mpie.de and r.ram@mpie.de

# Supplementary Figures

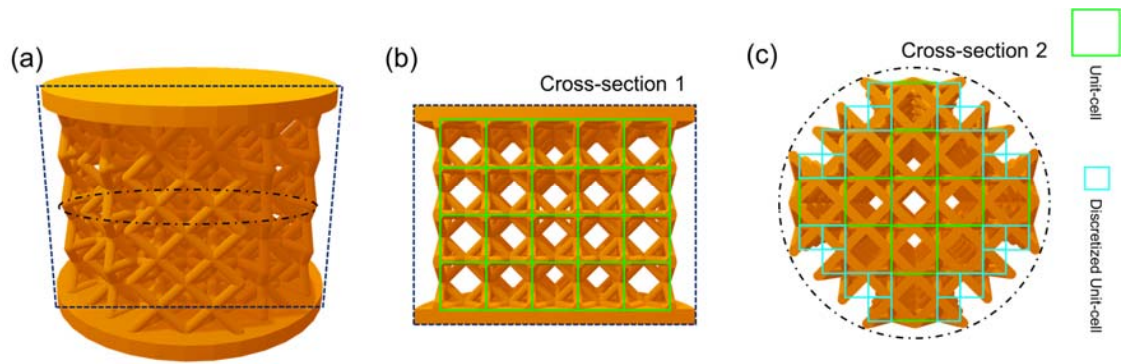

**Figure S1.** Computer-aided design of Cub structure. (a) Entire lattice structure, (b) Cross-section along building direction, and (c) Cross-section normal to building direction.

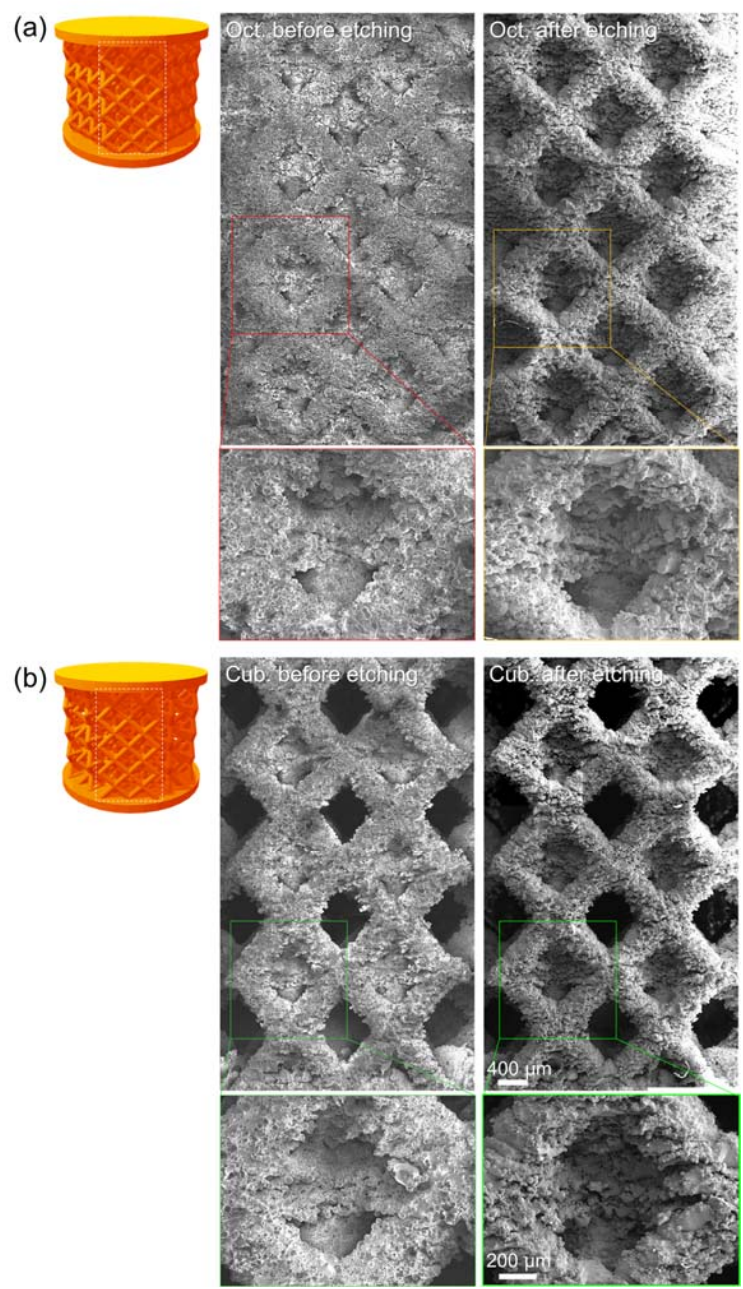

**Figure S2.** Chemical etching of pure Cu lattice structures. SEM images of (a) Oct and (b) Cub structures before and after chemical etching.

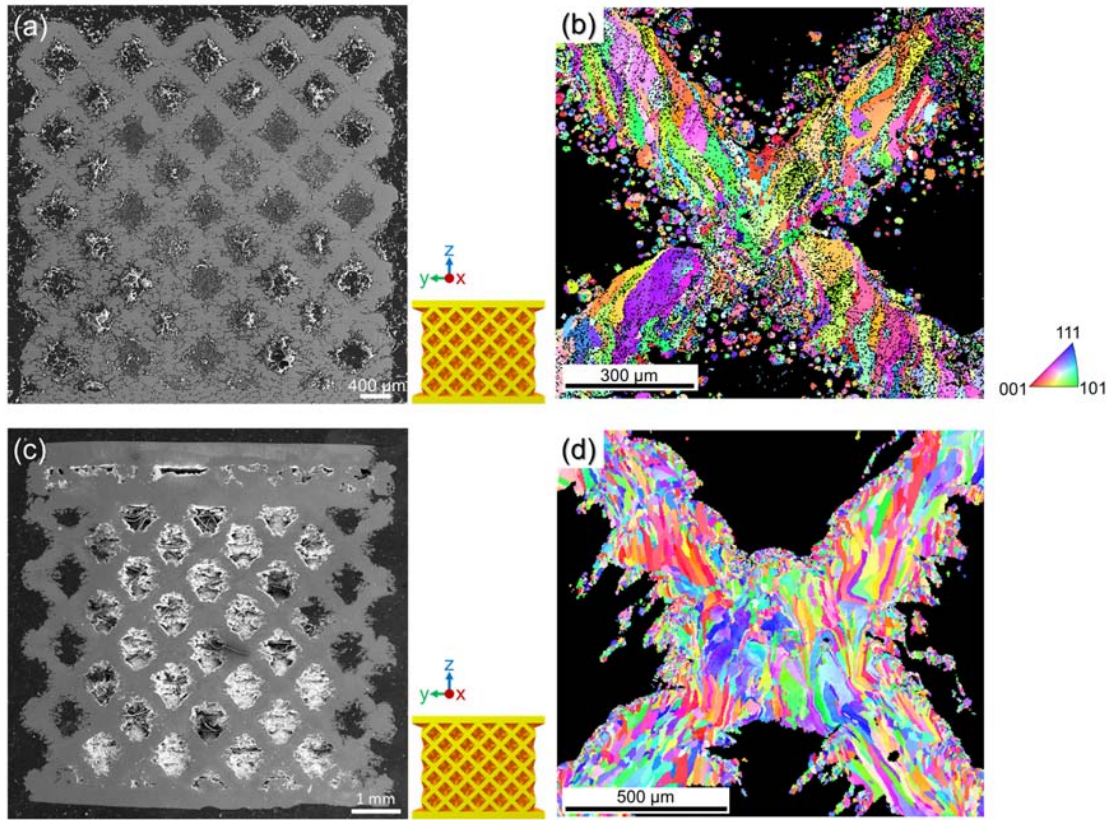

**Figure S3.** Comparison of microstructure of Oct structure fabricated by L-PBF with infra-red and green laser beams. SEM image and inverse pole figure map along building direction of Oct structure fabricated by (a, b) infra-red laser beam and (c, d) green laser beam.

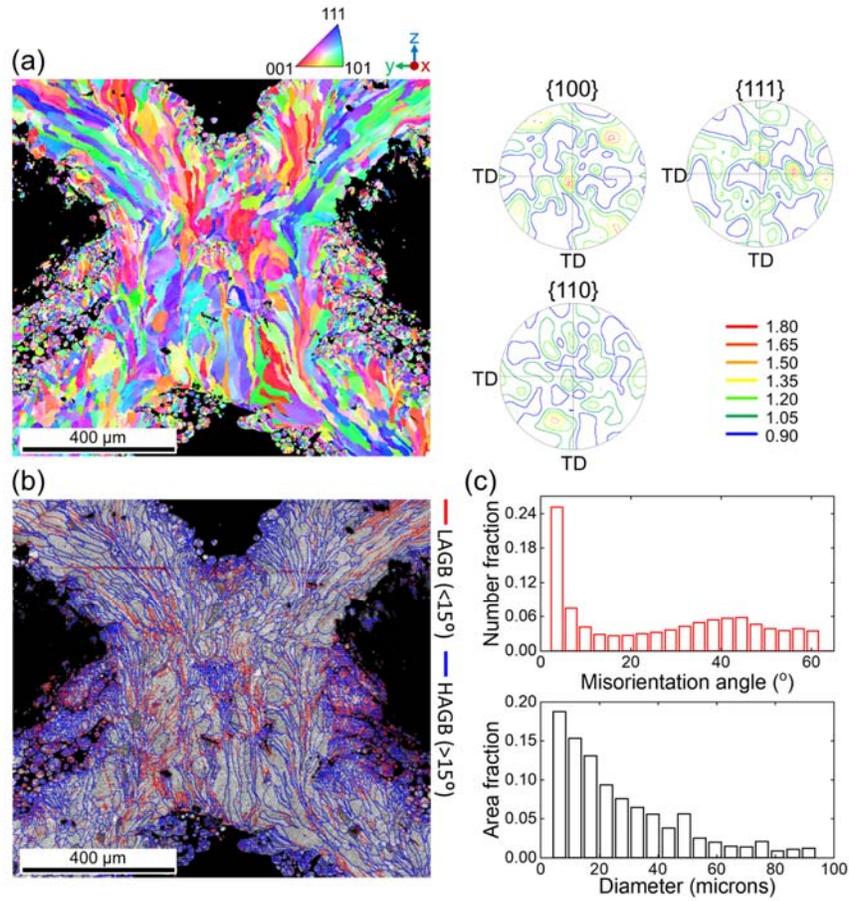

**Figure S4.** Microstructure of representative unit cross-section in Cub structure. (a) Inverse pole figure and corresponding pole figure maps along building direction. (b) Grain boundary map. (c) Misorientation angle distribution and grain size distribution plots.

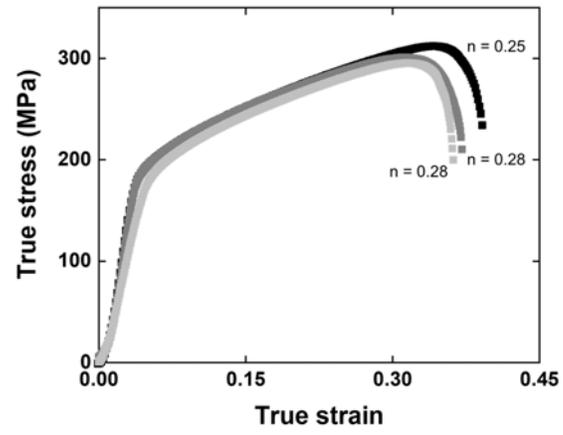

**Figure S5.** Quasi-static tensile test of dog bone pure Cu specimens fabricated by L-PBF with green laser beam.

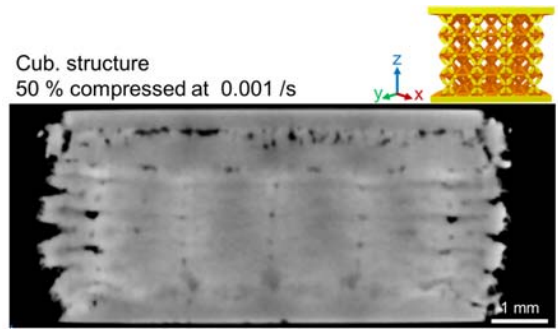

**Figure S6.** Cross-sectional microCT images of Cub structure compressed to 50 % at strain rate of 0.001 /s.

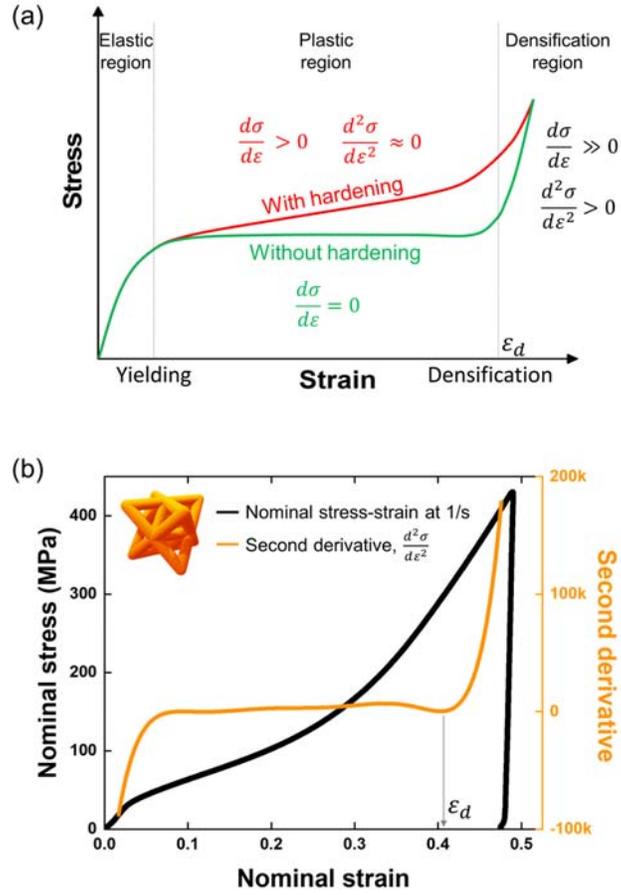

**Figure S7.** Determination of densifications strain of pure Cu lattice structure. (a) Schematic of compressive stress-strain curve with and without hardening behavior. (b) Nominal stress-strain curve of Oct structure and second derivative.

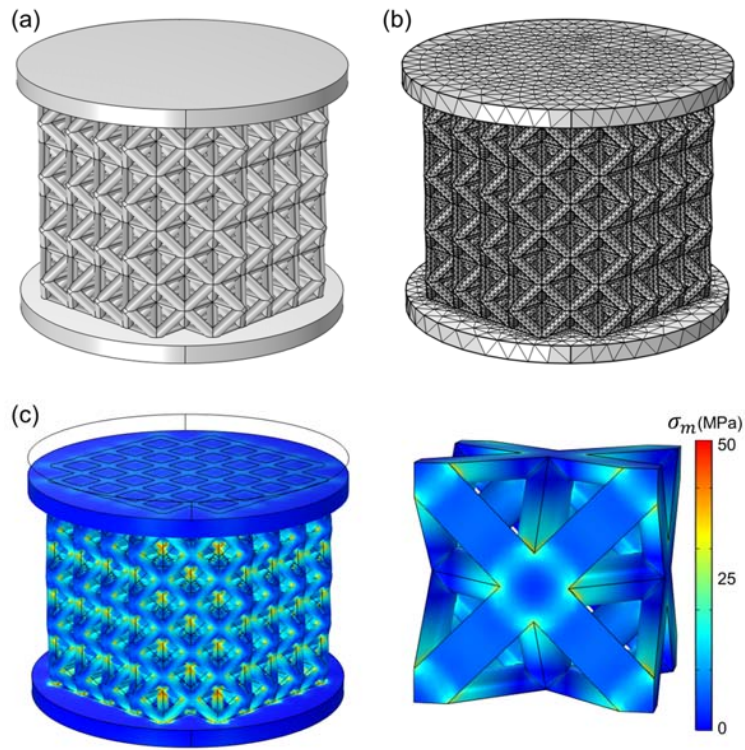

**Figure S8.** Finite element analysis of stress distribution in Oct structure under compression. (a, b) Model geometry and mesh distribution. (c) Mises stress distribution in Oct structure and unit-cell.